\documentclass[%
 reprint,
 amsmath,amssymb,
 aps,
]{revtex4-2}

\usepackage{graphicx}
\usepackage{dcolumn}
\usepackage{bm}
\usepackage{epstopdf}
\usepackage{color}

\begin{document}

\preprint{APS/123-QED}

\title{Magneto-optical trap reaction microscope for photoionization of cold strontium atoms}
\thanks{shenzhj@sari.ac.cn;\\ bing1.zhu@hsbc.com;\\ jiangyh3@shanghaitech.edu.cn}%

\author{Shushu Ruan,$^{1,3}$ Xinglong Yu,$^{1,3}$ Zhenjie Shen,$^{1,*}$ Xincheng Wang,$^{2}$ Jie Liu,$^{1,3}$ Zhixian Wu,$^{2}$ Canzhu Tan,$^{4}$ Peng Chen,$^{5,4}$ Tian-Min Yan,$^{1}$ Xueguang Ren,$^6$ Matthias Weidem\"uller,$^{7}$ Bing Zhu,$^{8,4,*}$ and Yuhai Jiang $^{2,1,3,*}$}
\affiliation{%
$^1$Shanghai Advanced Research Institute, Chinese Academy of Sciences, Shanghai 201210, China;\\
$^2$Center for Transformative Science and School of Physical Science and Technology, ShanghaiTech University, Shanghai 201210, China\\
$^3$University of Chinese Academy of Sciences, Beijing 100049, China\\
$^4$CAS Center For Excellence in Quantum Information and Quantum Physics,
University of Science and Technology of China, Hefei 230026, China\\
$^5$Advanced Materials and Center for Quantum and Science Technology, The Hong Kong University of Science and Technology, Guangzhou, China\\
$^6$School of physics, Xi'an Jiaotong University, Xian 710049, China\\
$^7$Physikalisches Institut, Universit\"at Heidelberg, Im Neuenheimer Feld 226, 69120 Heidelberg, Germany\\
$^8$HSBC Holdings Plc., Guangzhou 510510, China
}%


\date{\today}

\begin{abstract}
We developed a magneto-optical trap reaction microscope (MOTREMI) for strontium atoms by combining the multi-particle coincident detection with laser cooling technique. Present compact injection system enables the production of cold Sr atoms in three modes of 2D MOT, molasses and 3D MOT, providing adjustable densities and ratios of  the ground state $5s^2$ ($^1S_{0}$) and the excited states $5s5p$ $^{1}P_{1}$ and $^{3}P_{J}$ and so on. The target profiles for the temperature, the density and the size of the 3D MOT as well as the cold atomic flux in the 2D MOT model were characterized in details. Using present state-of-the-art setup, we successfully demonstrated the single photoionization of Sr atoms with molasses through absorption of a few 800-nm photons, where Sr$^+$ and $e$ were detected in coincidence and most of ionization channels were identified by considering photoelectron energy, laser-intensity dependence, and target characteristics. The best momentum resolution for coincident Sr$^+$ and $e$ along time-of-flight are achieved up to 0.12 a.u. and 0.02 a.u., respectively. Present  momentum and energy spectra of photoelectron ionized from the ground state and a few excited states illuminate unprecedentedly distinct structures manifesting prominent features related to multi-photon absorption. The complete vector momenta of electrons and recoil ions in coincidence open opportunities for further investigation of two-electron correlation dynamics and multi-electron effects in the multiple ionization of alkaline-earth atoms in the ultraviolet region.

\end{abstract}

\maketitle


\section{\label{sec:level1}INTRODUCTION}
Alkali metal and alkaline-earth metal play a crucial role in the study of atomic physics and quantum optics and have attracted increasing attention in the past few decades with the rapid advancements of cold atom technology. In addition to the promising potential applications in optical atomic clocks, quantum simulators and quantum computers, the combination of cold atoms  with electron/ion imaging technologies such as REMI-like (reaction spectroscope) and VMI-like (velocity-map imaging) has enabled investigations into  ion collision ionization \cite{IC1,IC2,IC3,IC4,IC5,IC6}, interatomic interaction of Rydberg atoms \cite{RIR}, penning ionization of cold metastable atoms and molecules \cite{PI1,PI2}, and cold molecular and chemical reactions \cite{CMCR1,CMCR2} as well as the strong field ionization \cite{SFI1,SFI2,SFI3,SFI4,SFI5,SFI6}. 

In the strong field ionization, except for well-known topics of above threshold ionization(ATI), tunneling ionization, nonsequential and sequential double ionization, the near-zero-energy structures of photoelectrons and attosecond high-order harmonic generation (HHG) and so on \cite{Review1,Review2,Review3,Review4,Review5}, in comparison with noble gases, alkali and alkaline-earth atoms with unique structural characteristics have distinctive features for investigations of HHG with higher harmonic frequencies \cite{MHHG1,MHHG2}, Kramers Henneberger (KH) atoms \cite{KH1}, the excitation polarization of atoms and the reconstruction of electron orbital wave functions \cite{wavefunction1,SFI4}, the double-excitation autoionization \cite{DEITBP1}, quantum interference in the single atom \cite{QI1}, and attosecond quantum simulator \cite{CAAQS1}. 

On the other hand, alkali metal and alkaline-earth elements are of relatively low ionization thresholds, making the commonly used adiabatic Keldysh parameter $\gamma$ unsuitable for characterizing their ionization regimes \cite{adiabaticparam1,SFI6}, despite its effectiveness with noble gas atoms. The subtle balance between the photon energy, the intensity of laser pulse, and the core potential makes it ambiguous to distinguish perturbative photon absorption and field-induced tunneling, implying a confusing but attractive interim between quantum and classical field ionization pictures. As a matter of fact, through the implementation of  MOTRIMS (magneto-optical trap recoil ion momentum spectroscopy \cite{Rb3}), we \cite{SFI6} recently observed the dominance of perturbative ionization even in the over-barrier regime where $\gg 1$. As the intensity increases, a transition from perturbative ionization to strongly perturbative ionization of rubidium atoms  becomes evident. Furthermore, recent theoretical studies \cite{shepherdele}  have revealed the presence of the shepherd electron effect in weakly bound electrons of alkali metals. This effect acts electron as a spectator, witnessing subsequent multiple ionization events and resulting in a double-knee structure and the emergence of a transient hollow atom in circularly polarized intense laser fields.

Compared to alkali metals, alkaline-earth atoms are of two-valence electrons outside a closed-subshell ionic core, complicated intermediate states of which are expected to involve in the reaction, particularly for atoms with higher atomic numbers \cite{HANum1,HANum2,Luc,HANum3,VMISr}. The presence of two-electron excited states creates a ladder-like distribution \cite{VMISr} near the ionization limits and within the continuum, presenting new challenges for the advancement of strong field theory. Moreover, this helium-like electron configuration offers an ideal system for investigating microscopic 3-body dynamics, such as double excitation autoionization \cite{DE1,DEITBP1,DE2} and the dynamics of two-electron strong correlations in double ionization \cite{Review4,He}, particularly when double excitation is involved.   Theoretical calculations are straightforward if a proper core potential is used \cite{theory1,theory2,theory3,theory4}. However, even for ion-yield measurements of alkaline-earth elements \cite{AEMEtheory1,AEMEtheory2} classical and semi-classical simulations \cite{classicalmodel1,classicalmodel2} have already exhibited distinct behaviors compared to noble gas atoms. This suggests the existence of novel mechanisms and poses challenges to current understandings. In the near future, there is great anticipation for kinematically complete measurements of double ionization, where two electrons and a doubly-ionized ion are detected in coincidence.

One major limitation in the experimental investigation of the aforementioned physical processes, particularly double ionization and multiple ionization, is the inadequate momentum resolution of recoil ions due to the gasification of alkaline-earth atoms and their large atomic masses. An advanced technology that addresses this challenge is the employment of MOTRIMS, as mentioned earlier. In the literature, MOTRIMS for several kinds of alkali metal atoms (Na \cite{IC1}, Rb \cite{Rb1,IC4,Rb2,Rb3}) has been developed, by which the experimental scopes about ion-atom collision and photoionization dynamics were successfully expanded into metal-element series. Following that efforts were made to improve MOTRIMS towards covering more sophisticated experiments in single-atom scale, one of which is kinematically complete experiment. The involving of reaction microscope presents a novel avenue for exploring the quantum properties of cold atoms. Overcoming the incompatibility between the magnetic fields respectively used for the formation of cold targets and electron momentum measurement, magneto-optical trap reaction microscope (MOTREMI), even all-optical trap combined with reaction microscope, has been realized for lithium \cite{SFI2,IC6,LiAOT2}. However, to the best of our knowledge, no such setup has been established for alkaline-earth atoms so far.

In this article, we report the first MOTREMI setup for strontium atoms to fill the blank of kinematically complete experiment about alkaline-earth-metal series. Strontium was chosen as a promising candidate due to a comprehensive assessment of technical feasibility and proposed physics, particularly its potential as a next-level optical frequency standard. Recent years have witnessed rapid advancements in laser cooling and frequency locking techniques for strontium, enabling the attainment of a stable compatibility between Sr-MOT target density and ultrahigh vacuum conditions. Beyond technological considerations, strontium exhibits numerous singly excited and doubly excited states in the ultraviolet (UV) waveband, allowing for coherent control of excited wave packets with infrared (IR) lasers and optical parametric amplifiers in laboratory. Moreover, the double ionization of cold strontium in the presence of either resonances or doubly excited states presents an intriguing topic for ongoing experiments.

The present paper is organized as follows. In the second section, the experimental setup is described in details along with the diagnostic results. In the third section, the results for the single ionization of cold strontium, as benchmark measurements, are shown and discussed. After a summary at the end of this paper, we give an outlook on future investigation.

\section{\label{sec:level2}MOTREMI for strontium atoms}
\begin{figure}
\centering
\includegraphics[width=8cm,height=7.5cm]{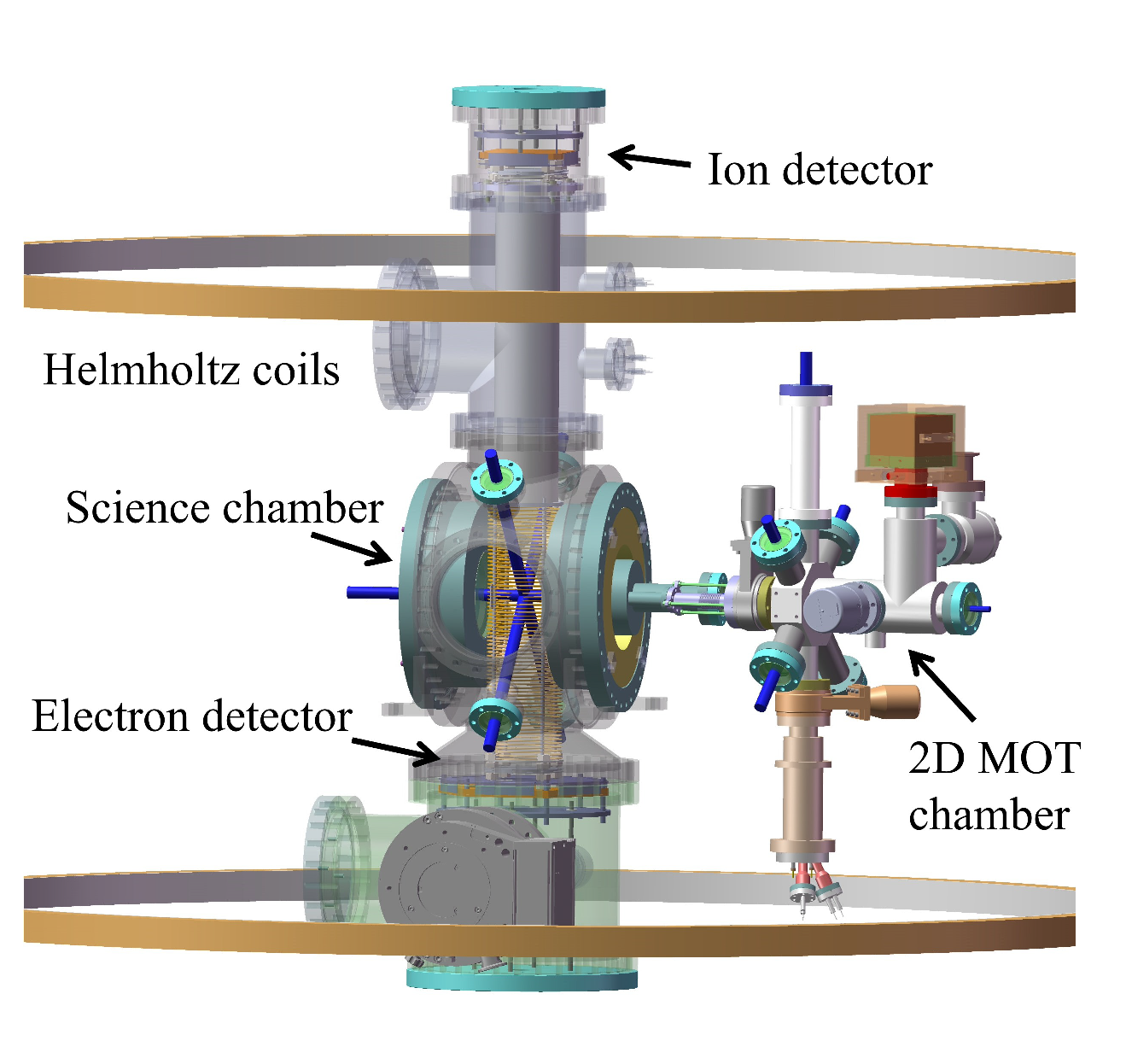}
\caption{\label{fig:1} The schematic diagram of Sr-MOTREMI. The apparatus structurally contains a 2D MOT chamber and a science chamber. The 2D MOT chamber is utilized to pre-cool hot strontium atoms and generate a stable cold atomic beam. The atomic beam will be cooled and trapped by 3D MOT in the center of science chamber, where the cold atoms are ionized by pulse laser. The reaction products in science chamber are collected by electron and ion detectors of a REMI.}
\end{figure}

The schematic diagram of the MOTREMI apparatus for strontium atoms is shown in Fig.~\ref{fig:1}, which mainly consists of a laser system, a sample injection system and a reaction microscope. The laser system provides frequency-locked laser beams essential for achieving laser cooling of strontium atoms. The sample injection system is realized for stable target sources with various modes for different experimental requirement. The reaction microscope will collect positive and negative charged fragments guided by a uniform electric field, and reconstruct relevant dynamical information of a reaction induced by a femtosecond laser pulse. Details of each component will be displayed in the following subsections. 

 The MOTREMI implementation involves a reaction microscope integrated with the magneto-optical trap technique for achieving exceptionally high momentum resolution of recoil ions. Due to the high melting point (769 $^{\circ}$C) of strontium solid, the heated atoms must undergo two-dimensional pre-cooling in the 2D MOT chamber to maintain high density and a cold beam in the science chamber.  Following propulsion by a weak laser beam (pushing beam), a 2D MOT beam will be directed to the center of the science chamber, where the atoms are further cooled and trapped by a 3D MOT. 2D MOT beam, molasses beam and 3D MOT target are ionized by pulse lasers and the charged fragments are collected by two position-sensitive detectors in a 4$\pi$ solid angle. The three-dimensional momentum of each particle is reconstructed from the time-of-fight (TOF) and hit positions on the detectors. 

\subsection{\label{sec:level2}Laser cooling system}
\begin{figure}
\centering
\includegraphics[width=9cm,height=7cm]{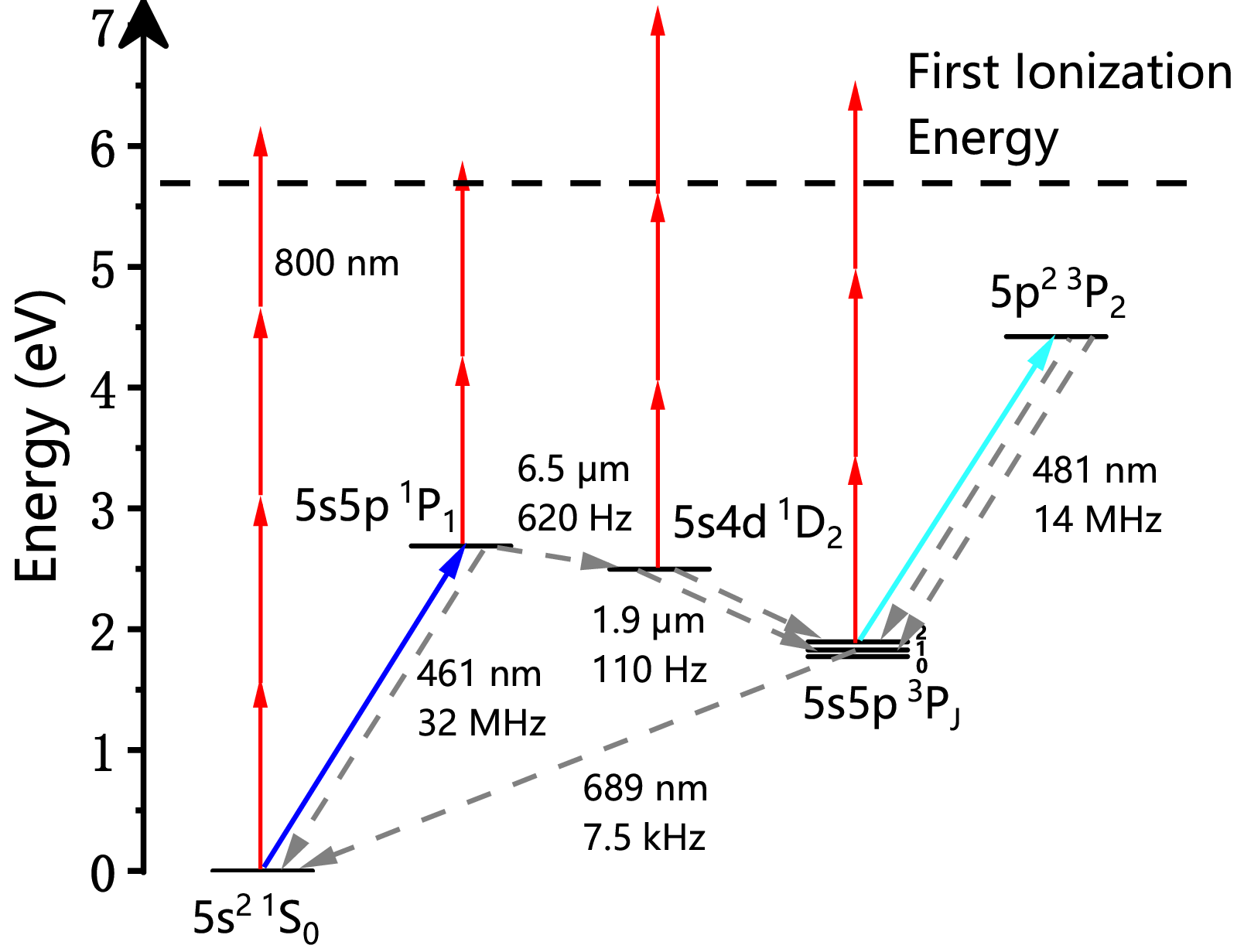}
\caption{\label{fig:energy level} Schematic diagram for relevant energy levels of $^{88}$Sr and laser-cooling transition pathways. The spectroscopic energy levels are taken from Ref. \cite{Srdata}. The natural linewidth and transition frequency are written in the diagram. Vertical arrows in red indicate possible ionization channels from the ground state and the excited states and arrow numbers are for the number of 800 nm photons required.}
\end{figure}

The schematic diagram of the relevant transitions for laser cooling and trapping of $^{88}$Sr is shown in Fig.~\ref{fig:energy level}. For laser cooling, the transition of $5s^2\ ^{1}S_{0}\leftrightarrow$ $5s5p\ ^{1}P_{1}$ at 461 nm is adopt. As the natural linewidth of $5s5p\ ^{1}P_{1}$ is $\Gamma$/2$\pi$ = 32 MHz, the Doppler temperature is 770 $\mu$K. Because of a weak spontaneous decay of $5s5p\ ^1P_{1}$ → $5s4d\ ^{1}D_{2}$ → $5s5p\ ^{3}P_{1,2}$, where $5s5p\ ^{3}P_2$ becomes the dark state with a lifetime on the order of 100 s, the transition $5s5p\ ^3P_2$ → $5p^2\ ^3P_2$ at 481 nm is chosen for repumping and then  $5p^2\ ^3P_2$ → $5s5p\ ^3P_1$ → $5s^2$. Repumping laser is not necessary for the 2D MOT since the decay ratio of transition $5s5p\ ^1P_{1}$ → $5s4d\ ^1D_{2}$ is only 1:50000 among the atoms on state $5s5p\ ^{1}P_{1}$.

\begin{figure}
\centering
\includegraphics[width=8cm,height=6.2cm]{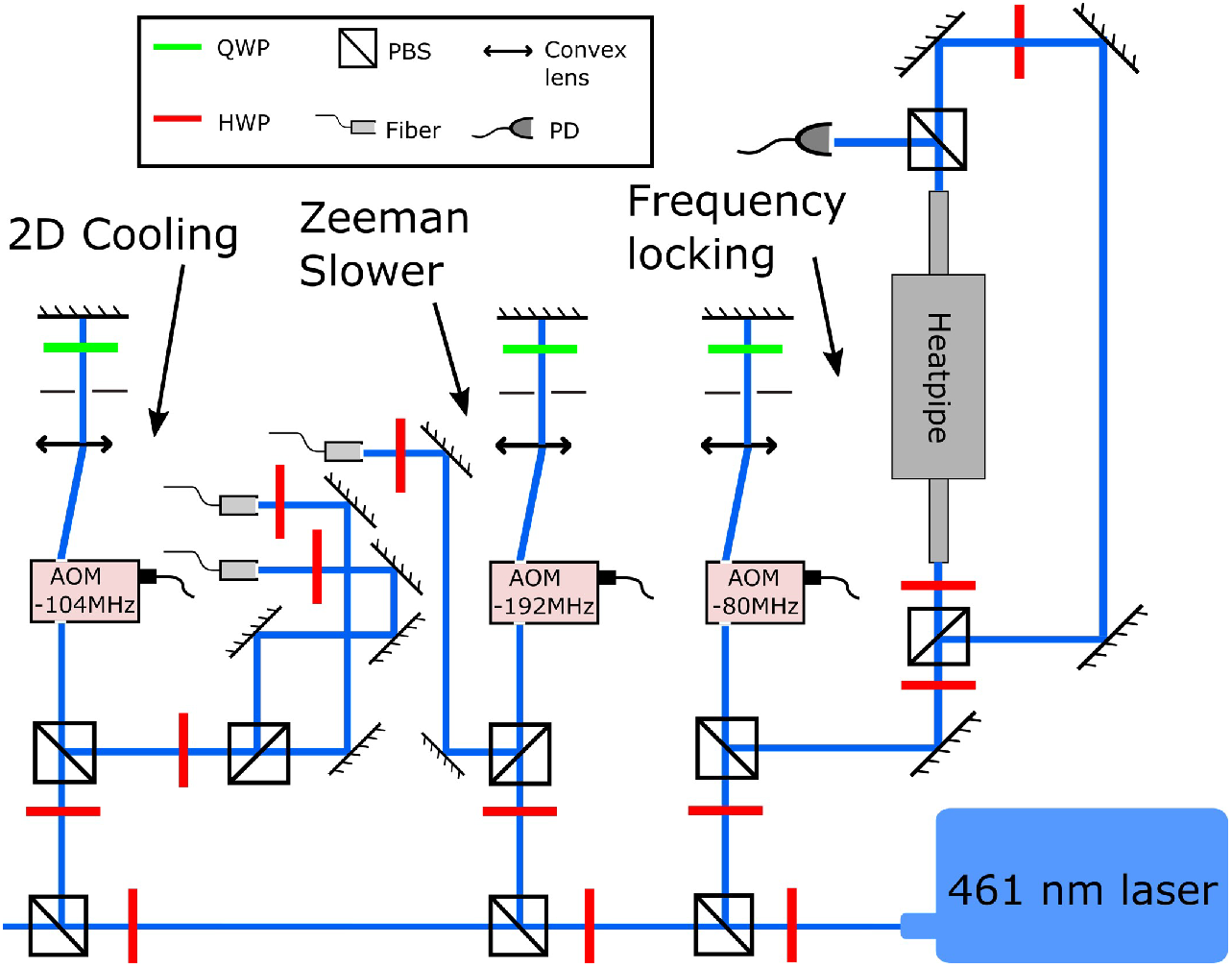}
\caption{\label{fig:optical path} Schematic diagram of part of the optical paths for laser cooling system. QWP:quarter-wave plate; HWP: half-wave plate; PBS: polarization beam splitter.}
\end{figure}

Laser cooling system in our apparatus comprises laser beams for Zeeman slower, 2D cooling, 3D cooling, pushing and frequency locking. Because of the similar optical setup of 3D cooling,  only part of the optical paths, i.e. Frequency locking, Zeeman slower and 2D cooling, is shown in Fig.~\ref{fig:optical path}. A commercial diode laser (Toptica DL Pro) is utilized to address the transition of $5s^2\ ^1S_{0}$ → $5s5p\ ^{1}P_{1}$ at 461 nm. In the 461-nm laser, a 922-nm seed laser with a few mW is amplified to $\sim$1.94 W by a tapered amplifier (Toptica TA). After passing through a frequency doubling cavity, a typical power of 600 mW (maximum exceeding 1 W) is delivered to the optical path. In the frequency locking module, the frequency stabilization is achieved by a saturated absorption spectroscopy system. The frequency of the 461-nm laser with few mW is shifted with 2×80 MHz by an acoustic-optical modulator (AOM) in a double-pass configuration, which is employed for all the other beams. The frequency-shifted laser (few mW) is divided by another PBS into a pump and a probe laser, both of which are directed towards a Sr heatpipe in opposite propagation directions. To generate an appropriate amount of atomic vapor, a metal strontium block inside the Sr heatpipe is heated to 320 $^{\circ}$C. The probe laser signal is detected by a photodiode (Thorlabs, PDA10A-EC) to obtain the saturated absorption spectrum. The frequency locking accuracy is approximately 1.6 MHz, which is significantly smaller than the natural linewidth of the laser cooling transition. Except for the frequency locking, all other laser beams are coupled into single-model polarization-maintaining fibers (Thorlabs, PM-S405-XP(PANDA)). Eventually, the achieved optical parameters of all the beams are summarized in Table.~\ref{tab:table1}.  

\begin{table}
\centering
\caption{\label{tab:table1}Configuration of laser cooling system. The table exhibits the power, frequency detuning ($\Delta$) and diameter of each laser beam of the laser system. The laser parameters are detected after being coupled into fibers.}
\begin{ruledtabular}
\begin{tabular}{cccc}
 &Power (mW)& $\Delta$ (MHz) & Diameter (mm)\\
\hline
2D cooling& 50 & -48 & 18\\
3D cooling& 10 & -30 & 12 \\
Zeeman slower& 50& -224 & 18 \\
Pushing& 0.33& 0 & 1.4 

\end{tabular}
\end{ruledtabular}
\end{table}

\subsection{\label{sec:level2}2D MOT chamber and 3D MOT science chamber}
\begin{figure}
\centering
\includegraphics[width=8cm,height=13cm]{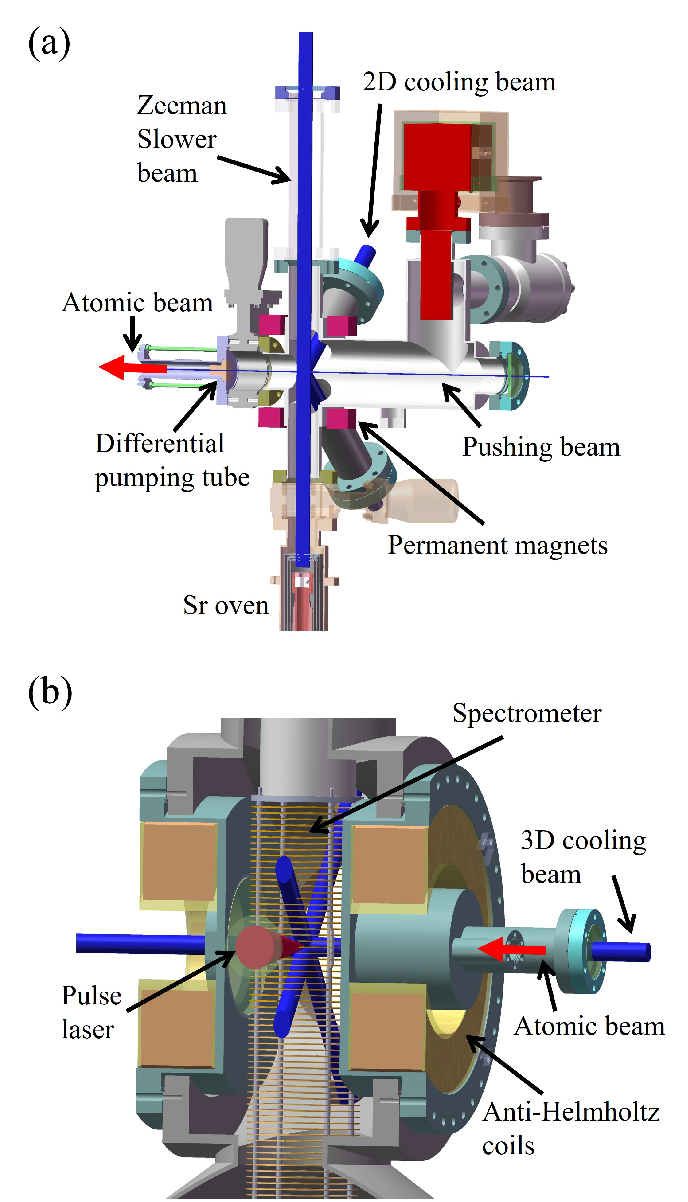}
\caption{\label{fig:2d} The zoom-in of 2D MOT chamber and 3D MOT science chamber. (a) The cutaway view of 2D MOT chamber. Hot Sr atoms evaporated from the Sr oven are firstly cooled in one-dimension by Zeeman slower and, then, cooled and trapped in the center of the chamber by 2D magneto-optical trap. The pushing beam will push the 2D MOT to form a cold atomic beam through a differential pumping tube into the science chamber. The magnetic field is generated by permanent magnets; (b) The cutaway view of 3D MOT science chamber. The 2D MOT cold atomic beam can be cooled and trapped by 3D magneto-optical trap in the center of science chamber. The cold targets will be ionized by a focused pulse laser, during which charged fragments are collected by electron and ion detector with the extraction of spectrometer.}
\end{figure}

The schematic diagrams of the 2D MOT chamber and 3D MOT science chamber are shown in Figs.~\ref{fig:2d}(a) and (b), respectively. Due to the high melting point, a metal strontium block is heated to the temperature of 450-550 $^{\circ}$C in the oven to generate a sufficient amount of atomic vapor ejecting into the 2D MOT chamber. For the following experiment measurements presented in this paper, the temperature is set to 480 $^{\circ}$C. In order to reduce the divergence of the atomic vapor, an array of dozens of microtubes are mounted at the head of the oven. The hot atoms are firstly decelerated by a laser beam propagating in the opposite direction to the hot atom beam, serving as a compact Zeeman slower \cite{IngoSr}. The decelerated atoms are subsequently cooled by two pairs of orthogonal retroreflected laser beams (2D cooling beam) and trapped at the center of the chamber, forming the 2D MOT. The required quadrupolar magnetic field of the 2D MOT is produced by four stacks of N35 neodymium (Nd2Fe14B) permanent magnets, which can also provide a magnetic-field gradient along the propagation direction of hot atomic vapor for the Zeeman slower. Each single piece of permanent magnets has three dimensions of 30 mm × 10 mm × 5 mm. To compensate for the positional shift of 2D MOT in the 2D chamber when the Helmholtz coils of reaction microscope are in operation, the permanent magnets are placed on the support of a three-dimensional transition stage for easy adjustment. Finally, the measured gradient of magnetic field in the trapping area along the direction of 2D cooling beams is approximately 47 G/cm.

The 2D MOT chamber is isolated from the science chamber by means of a differential pumping tube with a diameter of 2 mm and a length of 20 mm. A pushing beam is used to propel the 2D MOT to form a cold atomic beam which propagates through the differential pumping tube into the science chamber. The cold atomic beam is further cooled and trapped by three mutually perpendicular pairs of retro-reflected cooling beams (3D cooling beam). In Fig.~\ref{fig:2d}(b), it is noted that there is an angle of 12$^{\circ}$ between the atomic beam and one 3D cooling beam. The relevant parameters of the 3D cooling beam are presented in the Table.~\ref{tab:table1}. The quadrupole magnetic field generated by a pair of anti-Helmholtz coils are mounted coaxially outside the science chamber with a distance of 124 mm. Each coil has 112 turns with the number of radial layers of 14 and axial layers of 8, and the radius of the inner ring is 74 mm. At the current of 30 A, the coils can generate a magnetic field gradient of 32 G/cm along the axial direction. Additionally, the 481-nm repumping laser with a power of 10 mW and a 1/$e^2$ diameter of 1.4 mm is adopt to increase the atom number and prolong the lifetime of 3D MOT. 

Consequently, three modes of cold atomic targets, i.e. 2D MOT beam, molasses beam and 3D MOT, can be prepared. The 2D MOT atomic beam stems from the propulaion of cold atomic cloud from the 2D MOT chamber. The molasses beam is generated by six laser beams in the science chamber whereas the gradient magnetic field generated by anti-Helmholtz coils is absent. These targets with different atomic densities and state populations will be loaded for specific research purposes.

\subsection{\label{sec:level2}Spectrometer and detectors in science chamber}
We utilize a traditionally structured reaction microscope to measure the 3D momentum of charged fragments produced by ionization reactions. The working principle of the reaction microscope is summarized as that cold atomic targets are firstly ionized by a pulse laser in the center of the science chamber. Then the reaction products are extracted by a homogeneous electric field ($\sim$1.7 V/cm) generated by a spectrometer, at both ends of which the ion and electron detectors are located respectively. Eventually, relevant dynamic information will be reconstructed with the TOF and hit positions measured by the detectors. The spectrometer consists of 68 pieces of 1-mm thick stainless steel ring electrode with an inner diameter of 7.5 mm and an outer diameter of 10 mm. Each two adjacent electrodes with a distance of 4 mm are connected by a resistor of 100 k$\Omega$. In order to avoid eddy current when the magnetic field is switched off, each electrode needs to be cut off. The lengths of acceleration region of recoil ions and electrons are respectively 112.5 mm and 222.5 mm. Unique to recoil ions is the 430 mm long drift region. The few-V/cm electric field is adjusted to balance the detector resolution and the experimental detection efficiency. It should be noted that the assembled electrode needs to be sterically cut out of special holes for the 3D cooling beam and the femtosecond pulse laser. The pulse laser is focused by a concave spherical mirror mounted on a 3D manipulator.

With identical momenta, electrons move much faster than ions due to their small masses. In order to collect all electrons effectively, homogeneous magnetic field generated by a pair of Helmholtz coils is adopt. The electrons perform helical motions in the acceleration region, thereby constraining their trajectories. A pair of hollow copper coils, with an inner diameter of 6 mm and an outer diameter of 8 mm, is employed to create Helmholtz coils that utilize water cooling. The radius of each coil bundle is 83.5 cm. When a current of 10 A is applied, a magnetic field with a strength of 4.8 G can be generated.

The two-dimensional time and position sensitive micro-channel plate (MCP) with delay line anodes are employed on both ion and electron detectors. Both detectors employ a pair of 80-mm diameter MCPs. The ion detector utilizes square delay line anodes, while the electron detector uses hexagonal anodes to minimize dead time and enhance detection efficiency. The hit positions and flight times of charged fragments can be obtained through the data acquisition system, so as to reconstruct the physical parameters e.g., the initial momentum and energy. The detectors exhibit a time resolution of approximately 1 ns and a position resolution of 0.1 mm.

\section{\label{sec:level1}Characterization of sample injection system}

The target temperature and density have a significant impact on the momentum resolution and detection efficiency of our apparatus, ultimately affecting the feasibility of subsequent light-atom interaction experiments. Thus, it is imperative to conduct a diagnosis of our sample injection system to inform the design and analysis of our experimental investigation. To achieve this, we measure the flux of the 2D MOT beam and assess the characteristics of the 3D MOT, including the atomic number density and target temperature and so on. 

\subsection{\label{sec:level3}Cold atomic flux}
The flux of the 2D MOT beam is analyzed by detecting the time-varying fluorescence signal \cite{IngoSr}. A pair of retroreflected linearly polarized laser beam on resonance with the transition of $5s^2\ ^1S_{0}$ → $5s5p\ ^{1}P_{1}$ is applied to excite the cold atomic beam to produce fluorescence at the center of the science chamber, which has a distance of 410 mm from the center of 2D MOT. The excitation beams with a 1/$e^{2}$ radius of 3mm at a power of 12 mW are positioned on a pair of viewports used for 3D cooling laser beams, which are nearly perpendicular to the atomic beam. The fluorescence signal is collected using a lens with a focal length of 100 mm and a photodiode (Thorlabs, SM05PD1A). When the direction of the linear polarization of the excitation beam is perpendicular to the lens, the strongest signal is obtained, following the classical dipole radiation pattern. 

Suddenly cutting off the pushing beam, the temporal evolution of the fluorescence signal is measured, which determines the atomic flux. The atomic flux distribution on longitudinal velocity $v$ is given by

\begin{equation}
    \Phi(v)=-\frac{l}{m}\eta\frac{dU_{PD}(t)}{dt},\label{eq:1}
\end{equation}
where $l$, $\eta$ and $U_{PD}\left(t \right)$ are respectively the distance between the center of the 2D MOT chamber and the center of the science chamber, a calibration factor that combines the excitation efficiency and detection efficiency etc. of the whole fluorescence detector system, and the fluorescence voltage signal measured by the photodiode, respectively. Then the total atomic flux $\Phi_\text{total}$ can be calculated by the longitudinal velocity integration of Eq.(\ref{eq:1}).

\begin{figure}
\includegraphics[width=8.8cm,height=7cm]{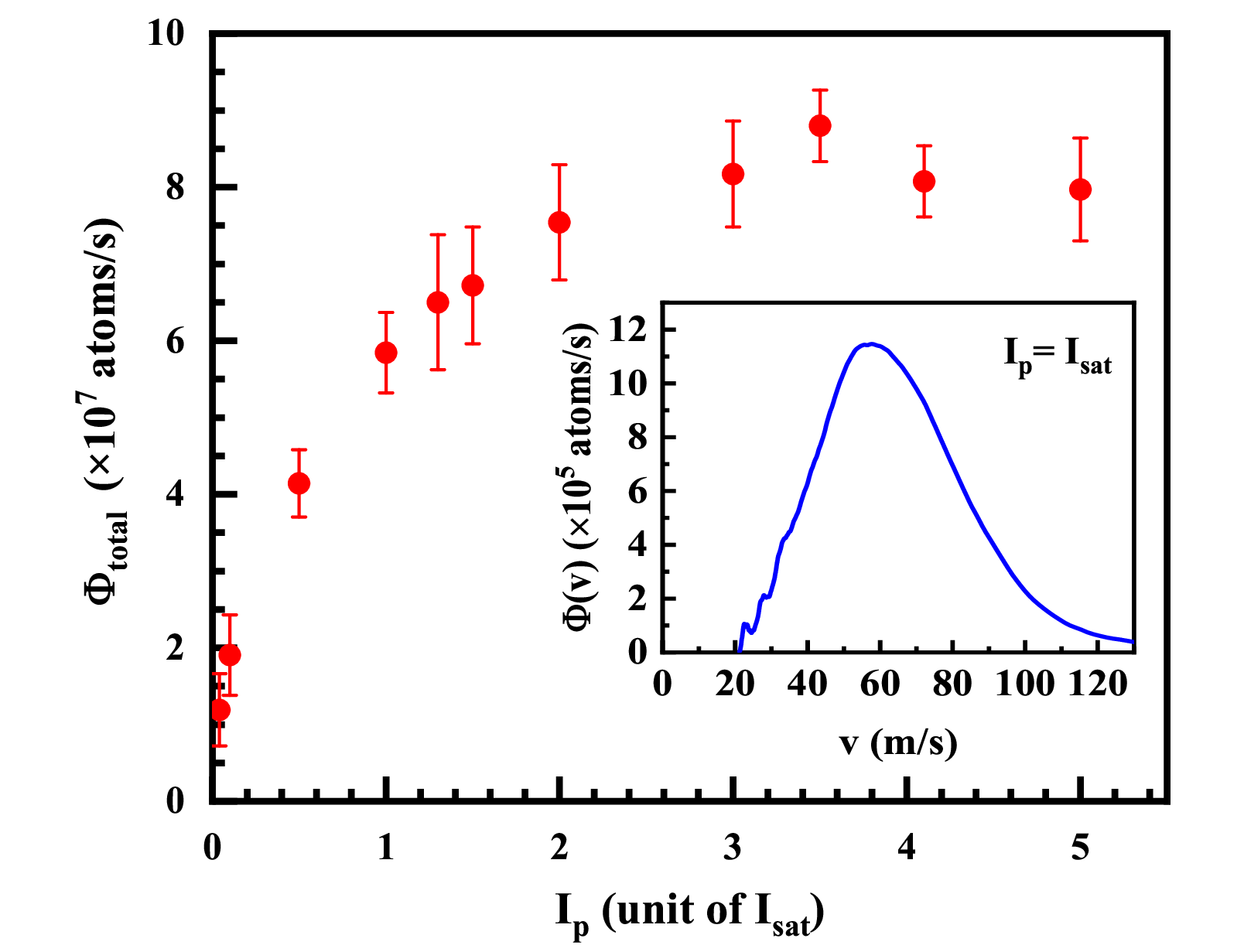}
\caption{\label{fig:flux} The total flux of 2D MOT beam versus the pushing laser intensity. The insets shows the longitudinal velocity distribution at pushing intensity of $I_p$ = $I_\text{sat}$.}
\end{figure}

The total atomic flux versus the laser intensity of the pushing beam is plotted in Fig.~\ref{fig:flux}, where $I_{p}$ is the peak intensity of the pushing beam. The insert of Fig.~\ref{fig:flux} shows the distribution of $\Phi(v)$ at $I_{p}$ = $I_\text{sat}$, with  a saturation density of $I_\text{sat}$=43 mW/cm$^2$ for the transition from $5s^2\ ^1S_{0}$ → $5s5p\ ^{1}P_{1}$. Here, the error bar of each point is derived from the calculated maximum inaccuracy of the measured data. It is clearly observed that $\Phi_\text{total}$ rises up to 8×10$^{7}$ atoms/s when $I_{p}$ increases to 2$I_\text{sat}$, and then remains relatively constant despite the further increase in $I_{p}$. 

For molasses beam, we have not characterized the target profiles in the same manner as the 2D MOT beam since its cooling lasers are continuously operational. Nonetheless, it was inferred that the density for the molasses MOT is several times higher than that of the 2D MOT, as deduced from the ionization rate.

\begin{figure}
\centering
\includegraphics[width=7cm,height=8cm]{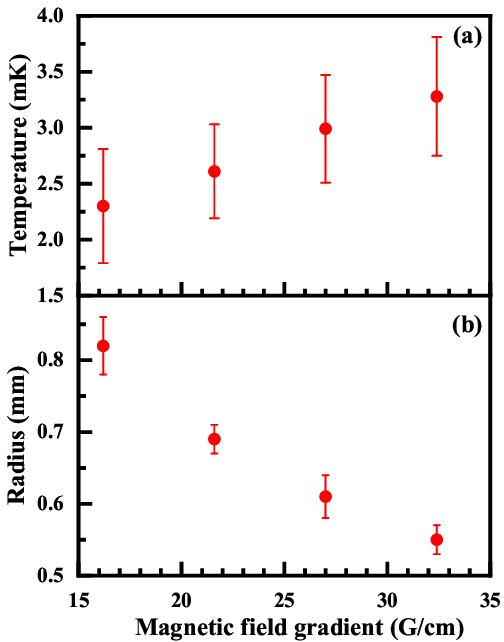}
\caption{\label{fig:3D MOT} Experimental magnetic field dependence of temperature and size of 3D MOT. The temperature (a) and radius (b) of 3D MOT are plotted as a function of the magnetic field gradient.}
\end{figure}

\begin{figure*}
\centering
\includegraphics[width=18cm,height=6cm]{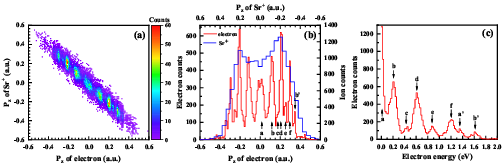}
\caption{\label{fig:momentum coincidence} The coincident measurement of single photoionization of molasses at the IR laser intensity $I$ = 3 TW/cm$^2$. (a) The momentum coincidence of Sr$^+$ and electron along $z$ direction. The distribution is obtained by slicing the momentum of electron with $|P_x|\le$ 0.06 a.u. and $|P_y|\le$ 0.06 a.u.. (b) The momentum spectra of electron (red line) and Sr$^+$ (blue line) along $z$ direction. The multi-peak structures in momentum spectrum of electron are marked as $a$-$f$ and $a'$-$b'$. (c) The photoelectron energy spectrum are calculated by the momentum of electron in $P_z$ with $|P_x|\le$ 0.06 a.u. and $|P_y|\le$ 0.06 a.u.. Each peak is also marked corresponding to the structures in the momentum spectrum of electron.}
\end{figure*}

\subsection{\label{sec:level3}3D MOT characterizations}
The temperature of the 3D MOT is obtained by measuring its free expansion image of the atomic cloud using absorption imaging method. A few-$\mu$W excitation beam is coupled into the same fiber with one 3D cooling laser beam. The switching of the imaging beam and all 3D cooling beams are controlled sequentially by AOMs. The excitation beam with 3D MOT and without 3D MOT at different times are recorded as the data image and the background image by a CMOS camera, respectively. The 3D MOT image is determined by the subtraction of the above two kinds of image. The standard deviations $\sigma_{x}$ and $\sigma_{y}$ are extracted with a two-dimensional Gaussian fitting on the 3D MOT image assuming the atomic spatial Gaussian distribution. Here $x$ and $y$ are any two directions perpendicular to each other. The standard deviation on free expansion time $t$ is determined by $\sigma^2\left(t \right) = \sigma^2_0+\frac{k_BT}{m}t^2$, where $\sigma_0$, $k_{B}$, $T$, $m$ are  the standard deviation at the initial moment, the Boltzmann constant, the atomic cloud temperature and the strontium atomic mass, respectively. With linear fitting, the temperature is figured out by the slope. When the magnetic field gradient is 32.4 G/cm, the temperature along the direction of $x$ axis and $y$ axis are respectively 3.0 mK and 3.6 mK, corresponding to the arithmetic mean value of the two temperature $T_\text{mean}$ = 3.3 mK. The temperature and size of 3D MOT on the magnetic field gradient are shown as a function of the magnetic field in Fig.~\ref{fig:3D MOT}. With the increase of the magnetic field gradient, the temperature increases from 2.3 mK to 3.3 mK and the radius of the MOT decreases from 0.82 mm to 0.55 mm.

The 3D MOT density can also be characterized by the absorption imaging method. In addition to the absorption image (with excitation beam and 3D MOT) and the reference image (with excitation beam but without 3D MOT), the background image (without both) is also need to be recorded. The distribution of the optical density is determined by following the Lambert-Beer law and the atom number is estimated to 3.7×10$^6$ atoms. Assuming that the 3D MOT is a sphere with radius of 0.55 mm (seen in Fig.~\ref{fig:3D MOT}), the average atom density is estimated to be 5.3×10$^9$ atoms/cm$^3$.

\section{\label{sec:level2}Photoionization of Strontium}

\subsection{\label{sec:level2}Momentum spectra of electron and ion in coincidence}
The single photoionization process is employed to characterize the momentum resolution of photoelectrons and photoions in the current setup. A laser pulse delivered by Ti:sapphire femtosecond laser with 800-nm wavelength, 35-fs pulse duration and 1-kHz repetition is utilized to ionize cold Sr atoms. The laser intensity of about 3 TW/cm$^2$ used was estimated by comparing the stabilization appearance of the peak induced by four-photon ionization of the ground state at $I$ $\approx$ 6 TW/cm$^2$ in Ref. \cite{VMISr}. This calibrated intensity is consistent with our previous calibration method detailed in Ref. \cite{SFI3}, employing similar IR laser setups for both measurements. In this study, a molasses target is chosen instead of a 3D MOT, allowing for full electron detection without the need to switch the anti-Helmholtz coils, thereby improving collection efficiency. Meanwhile, in contrast to 2D MOT the molasses targets contain a certain fraction of excited states, which can help us to identify photoelectron spectra. On the other hand, the 481-nm repumping laser is not applied since we found that it might result in more initial states and make part of the spectra strongly overlapped. 

 In the coordinate system of our setup, the directions of TOF, pulse laser propagation, and perpendicular to both (cold atomic beam) are defined as  $z$, $y$, and $x$, respectively. The laser polarization is along the $z$ direction (TOF direction). In Fig.~\ref{fig:momentum coincidence}(a) by slicing $P_x$ and $P_y$ of electron with 0.06 a.u., we exhibit the coincident momenta of Sr$^+$ and electron along the TOF. The data reveals a clear anti-diagonal line, signifying the momentum conservation between the electron and the ion in the TOF direction. The varying structures on these data corresponding to different energies of electron and recoil-ion indicate the existence of photoionization of a few of initial states. The corresponding momentum spectra $P_z$ of electron and Sr$^+$ projected into the TOF direction are shown in Fig.~\ref{fig:momentum coincidence}(b), where rich multi-peak structures marked as $a$-$f$, $a'$ and $b'$ can be found in the momentum distribution $P_z$ of electron. With the multi-peak fitting, the momentum resolution of electron up to 0.02 a.u. can be achieved at the full width at half maximum (FWHM). Due to the strong overlapping of those peaks the momentum spectra of Sr$^+$ displayed a broad peak shown in Fig.~\ref{fig:momentum coincidence}(b), which closely resembles electron momentum spectra $P_z$ with ion's resolution. With selecting the peak at $\pm$0.2 a.u. in Fig.~\ref{fig:momentum coincidence}(a), we extract the best momentum resolution of 0.12 a.u. at FWHM for ion along $P_z$. In Fig.~\ref{fig:momentum coincidence}(c), the energy spectra of photoelectron calculated from spectra in Fig.~\ref{fig:momentum coincidence}(b) are plotted, revealing an additional small peak at an electron energy of 1.35 eV. 

\subsection{\label{sec:level2} Single photoionization of Sr in the ground state and excited states}
In Fig.~\ref{fig:energy level}, we draw possible ionization pathways indicated by vertical uparrows. The length and quantity of arrows present 800-nm photon energy and the number of photons required energetically, respectively. These ionization pathways are described as
\begin{eqnarray}
    &\rm{Sr}&(5s^2 \ ^1S_0)+4 \hbar \omega \rightarrow {\rm{Sr}}^+ +e (0.51\,\rm{eV}),\label{eq:3}\\
    &\rm{Sr}&(5s5p\ ^1P_1)+2 \hbar\omega \rightarrow {\rm{Sr}}^+ +e (0.1\,\rm{eV}),\label{eq:4}\\
    &\rm{Sr}&(5s4d \ ^1D_2)+3 \hbar\omega \rightarrow {\rm{Sr}}^+ +e (1.45\,\rm{eV}),\label{eq:5}\\
    &\rm{Sr}&(5s5p \ ^3P_2)+3 \hbar \omega \rightarrow {\rm{Sr}}^+ +e (0.75\,\rm{eV}),\label{eq:6}
\end{eqnarray}
where $\hbar\omega$ = 1.55 eV. $5s^2\ ^1S_0$ and $5s5p\ ^1P_1$ are the ground state and the excited state pumped by 461 nm cooling laser. $5s4d\ ^1D_2$ and $5s5p\ ^3P_{1,2}$ are populated via decay of $5s5p\ ^1P_1$ and $5s5p\ ^3P_2$ is a dark state with 100 seconds lifetime. Except for photoelectron energies written in brackets (see Eqs.(\ref{eq:3})-(\ref{eq:6})), one notes that these ionization channels are required energetically to absorb 2$-$4 photons. The relative ionization probabilities sensitively depend on laser intensity $I$ by $\sim I^n$ ($n$ for the number of photons absorbed), which can be considered to be the second  basis for analyzing photoelectron spectra. Additionally, the power of the cooling laser can modify the proportion of the ground state and the excited states, assisting in identifying the ionization processes of these states.  

Considering only electron energy spectra complicates the determination of ionization pathways due to the ponderomotive potential $U_p$ shifting the kinetic energies of electrons.  Therefore, firstly, using molasses MOT and 2D MOT, we found that $a$, $a'$ and $c$-$f$ peaks, shown in Fig.~\ref{fig:momentum coincidence}(c), result from the excited states and $b$ and $b'$ peaks are from the ground state. Meanwhile, by laser intensity dependence not shown here, $a$, $a'$ and $c$-$f$ channels present less numbers of absorbed photon than that of $b$ and $b'$ channels. On the basis of discussions above, we can clearly conclude that electrons with near zero energy marked as the peak $a$ are generated from two-photon ionization of the excited state $5s5p\ ^{1}P_{1}$ (Eq.(\ref{eq:4})).  Electrons of peak $b$ located at about 0.2 eV are nominated to the ionization of the ground state $5s^2\ ^1S_{0}$ by absorption of four photons, where momentum distribution, not shown in this article, demonstrates one structure towards $5s\varepsilon d$ continuum (D state). The features are similar with that reported in Ref. \cite{VMISr}. Electron energy variation of ~0.3 eV in comparison with theoretical value given in Eq.~(\ref{eq:3}) results from U$_p$ shift. When target densities of 2D and molasses are changed, we found that the ionization ratios of peak $b$ and peak $b'$ at 1.62 eV as well as peak $a$ and peak $a'$ at 1.35 eV remain constant, which means that these two peaks stem from the same initial state. Considering additionally their kinetic energies and laser intensity dependence,  $b'$ and $a'$ peaks are the first ATIs of five-photon ionization of the ground state and three-photon ionization of the excited $^1P_1$ state, respectively. Electrons of peaks $d$ at 0.62 eV and $f$ at 1.2 eV are inferred to three-photon ionizations of excited states $5s5p\ ^3P_2$ and $5s4d\ ^{1}D_{2}$ described in Eqs.~(\ref{eq:6}) and (\ref{eq:5}), respectively. Due to the long lifetime and three-photon absorption photoionization of $5s5p\ ^3P_2$ is always pronounced even if 2D targets are used. This indicates that $5s5p\ ^3P_2$ state can be populated in the 2D chamber where cooling lasers are always on, and then they flow into the science chamber. Small peak $e$ is of dependence of three-photon absorption so that it might originate from ionization of $5s4d\ ^{3}D_{1,2}$ state, not shown in Fig.~\ref{fig:energy level}, although the decay pathway $5s5p\ ^1P_1 \rightarrow 5s4d\ ^{3}D_{1,2}$ is assumed to be much weaker than $5s5p\ ^1P_1 \rightarrow 5s4d\ ^{1}D_{2}$ \cite{state3D}. 

 Unfortunately, we cannot clearly identify another small peak $c$ according to energy levels, where it presents three-photon laser intensity dependence and meanwhile its kinetic energy is independent of laser intensity. Therefore, we can excluded the contributions of ionization with excitation, where four photons are needed at least to reach the lowest excited Sr$^+(4p^64d$) state. One of possibilities might result from other $J$ quantum number states, for instance $5s5p\ ^3P_0$ state.  Energy level splittings ($\sim$0.1 eV) of $5s5p\ ^3P_2$ and $5s5p\ ^3P_0$ states are much smaller than observations. However, $U_p$ effects to different states could lead to different shifts, which might increase separations of photoelectron energies. 
 
Relevant momentum distributions for states discussed above illuminates amazing landscapes manifesting prominent features induced by absorption of a few photons, which contains information of different initial and final angular momenta. The amplitudes and phases of different ionization channels reflect intrinsic photoionization dynamics.  Detailed studies for laser intensity-dependent and state-dependent momentum distributions will be addressed in details in the ongoing article. 

\section{\label{sec:level1}SUMMARY}

We report in details the first complement of the Sr-MOTREMI setup with the combination of the multi-particle coincident detection and laser cooling techniques, which   consists of frequency-locked laser system, cold atom injection system and detection system called a reaction microscope. These components are responsible for providing the necessary laser beams for MOT formation, stable and high flux atoms, and highly effective collection of charged particles, respectively. In present experimental configuration, the 2D MOT beam flux is up to 8×10$^{7}$ atoms/s, several times lower than the density of molasses beam estimated by counting ionization rate. The density of 3D MOT with a radius of 0.55 mm and the temperature of 3.3 mK is about 5.3×10$^{9}$ atoms/cm$^3$ with the magnetic field gradient of 32 G/cm. As a proof-of-principle experiment, we demonstrated the single photoionization of Sr atoms with a molasses target by absorption of a few 800-nm photons, where Sr$^+$ and $e$ were detected in coincidence.  The best momentum resolution along time-of-flight for coincident Sr$^+$ and $e$ are achieved up to 0.12 a.u. and 0.02 a.u., respectively. Multiple ionization channels induced by absorption of a few photons are discussed and identified by analyzing photoelectron energy and momentum spectra. With the current state-of-the-art technology, we are able to observe photoelectron momentum and energy spectra ionized from the ground state and the excited states, revealing unprecedentedly rich structures, which presents prominent features related to multi-photon absorption initiating a novel way for study of photoionization dynamics.

To our knowledge, this is the first electron-ion coincident measurement of an alkaline earth element, with which the population of various excited states is under control and photoionization of the excited states is within reach. The single ionization experiment serves as a foundation for investigating multiple ionization processes. Beyond the single ionization, for example, the double ionization of strontium, helium-like atom, promises an interesting process that is anticipated to exhibit distinctive two-electron correlation dynamics. 

\acknowledgments
 This work is supported by the National Natural Science Foundation of China (Grants Nos. 11827806 and 12334011) and the National Key Research and Development Program of China (Grants No. 2022YFA1604302). We are grateful to Ingo Nosske and Fachao Hu for their constructive discussions.

\nocite{*}

\bibliography{Sr-MOTReMi}

\end{document}